\documentstyle[jkas]{article}

\beginpage{7}
\endpage{17}
\year{2012}\volume{45}\month{February}\issueno{1}
\runningauthor {C. CHOI ET AL.}
\runningtitle{A $Y$-BAND LOOK OF THE SKY}

\month{February} \year{2012} \volume{45}
\beginpage{7}\endpage{17}
\date{Received November 16, 2011; Revised December 27, 2011; Accepted January 2, 2012}

\begin{document}
\title{A $Y$-BAND LOOK OF THE SKY WITH 1-M CLASS TELESCOPES}
\author{Changsu Choi$^{1}$, Myungshin Im$^1$, Yiseul Jeon$^1$, and Mansur Ibrahimov$^2$}
\address{$^1$ CEOU/Astronomy Program, Department of Physics and Astronomy, \\Seoul National University,
Seoul 151-742, Korea\\ {\it E-mail : changsu@astro.snu.ac.kr, mim@astro.snu.ac.kr}}
\address{$^2$ Ulugh Beg Astronomical Institute, Tashkent, Uzbekistan}
%\\ {\it E-mail : mansur@astrin.uz}}
\address{\normalsize{\it (Received November 16, 2011; Revised December 27, 2011; Accepted January 2, 2012)}}
\offprints{M. Im}
%--------------------------------------------------------------------
%Im I shortened the abstract, and removed sentences which are
%Im unncessary.
\abstract{ $Y$-band is a broad passband that is centered at $\sim$1 $\mu m$.
 It is becoming a new, popular window for extragalactic study especially for observations
 of red objects thanks to recent CCD technology developments.
 In order to better understand the general characteristics of objects in $Y$-band,
 and to investigate the promise of $Y$-band observations with small telescopes,
 we carried out imaging observations of several extragalactic fields, brown dwarfs, and
 high redshift quasars with $Y$-band filter at
 the Mt. Lemmon Optical Astronomy Observatory and the Maidanak observatory.
 From our observations, we constrain the bright end of the galaxy
 and the stellar number counts in $Y$-band. We also test the usefulness of high redshift
 quasar ($z>$6) selection via $i-z-Y$ color-color diagram,
 to demonstrate that the $i-z-Y$ color-color diagram is effective for the selection of
 high redshift quasars even with a conventional optical CCD camera installed at a 1-m class telescope.
  }

\maketitle

%--------------------------------------------------------------------
\section{INTRODUCTION}

 Recently, there has been an enormous progress in the studies of Near-InfraRed (NIR) bright sources.
 Large area surveys such as the Sloan Digital Sky Survey (SDSS; \citealt{Abazajian08}), the Two Micron All Sky Survey (2MASS; \citealt{Skrutskie06}), the 2 degree Fields Galaxy Redshift Survey (2dFGRS; Colless et al. 2001), the UKIRT Infrared Deep Sky Survey (UKIDSS; \citealt{Lawrence07}) and deep imaging observations from space and the ground led to the discovery of rare, interesting objects such as quasars at the epoch near re-ionization of the universe (\citealt{Fan00}), the understanding of distant massive galaxies and clusters in formation (\citealt{Kang09}; \citealt{Shim07}), and the discovery of many brown dwarfs such as T/L-dwarfs with surface temperature around 1000 K or below (\citealt{Chiu07}).
 Many of these studies start with extensive ground-based or space-based imaging observations in optical (from $u$-band to $z$-band, or 0.3551 $\mu m$ $-$ 0.8931 $\mu m$), or in NIR ($J$-band to $K$-band, or 1.27 $\mu m$ $-$ 2.2 $\mu m$) over a large-area in the sky, or very deeply in small fields. Spectroscopic observations have been carried out to unveil the true nature of interesting astronomical sources selected from the multi-wavelength imaging data.
These previous studies have provided valuable multi-wavelength data-sets, but there still remains a gap in wavelength coverage between the optical and the NIR, namely 1 $\mu m$ wavelength region, which is much less explored compared to
 the other wavelengths.

 The wavelength region of $Y$-band is centered at $\sim$1 $\mu m$ with a spread of
 about 0.1 $\mu m$ (\citealt{Hillenbrand02}).
 The $Y$-band is located at a relatively clean atmospheric window between NIR and optical bands.
 Fig. 1 shows a typical $Y$-band filter transmission curve compared to other red, optical bands and
 $J$-band. The name $Y$ is dubbed to distinguish it from $z$ and $Z$ filters which cover the wavelength close to $Y$.
% The Hubble Space Telescope (\textit{HST}) also adopted filters which are similar to $Y$-band.
 Traditionally, this wavelength has been neglected mainly due to the lack of detectors
 that can cover this wavelength in a cost-effective way.
 But the situation is changing with the emergence of deep depletion CCD chips that boast excellent
 sensitivity at 1.0 $\mu$m (QE $\sim$ 25 $-$ 50\%; e.g., Park et al. 2011),
 compared to those of the traditional back-illuminated,
 thinned CCDs (QE $\sim$ 5 $-$ 10\%; e.g., Lee et al. 2010; Im et al. 2010). These CCDs are cost-effective
 compared to expensive NIR detectors made of InSb or HgCdTe,
 making observations with $Y$-band more popular than ever.

\begin{figure}[!t]
\centering \epsfxsize=8cm
\epsfbox{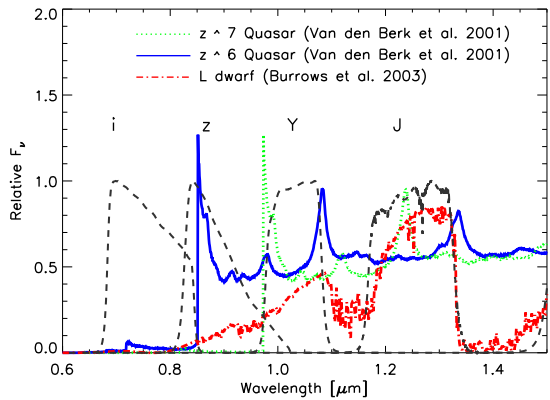}
\caption{Spectral Energy Distributions (SEDs) of a $z \sim$6 quasar
 (the solid blue line), $z \sim$7 quasar
 (the dotted green line) and L dwarf star (the red dot-dashed line).
 The black dashed lines represent the relative transmission efficiencies of
 $i,z,Y$, and $J$ filters. These types of objects have very red
 $i-z$ or $z-Y$ colors,
 and the addition of $Y$ or $J$-band helps
 to distinguish high redshift quasars from brown dwarfs.}
\label{fig1}
\end{figure}

  Early works with $Y$-band were mainly limited to studies of cool and
 low mass stars like brown dwarfs.
  A pioneering work of Hillenbrand et al. (2002) attempted to provide photometric calibration
 for $Y$-band, and demonstrated the usefulness of $Y$-band for studying cool stars.
% In their study, they found $Y$-band photometric informations, the slope of interstellar reddening
% vector within $Y$-band A$_{Y}$=0.38 $\times$ A$_{V}$.
%  They found  deep molecular absorption features in NIR spectra of extremely cool star like M and L
% dwarfs show blue-ward in L dwarf stars than T dwarf in $Y-H$,
% $Y-K$ colors.
  $Y$-band is useful for the identification of low-mass stars especially at young ages, and can
 distinguish low mass stars from high mass stars even in the galactic plane and
 in star-forming regions without too much interruption due reddening.

  Also, extragalactic research can benefit from $Y$-band imaging greatly.
  These days, the forefront of the extragalactic astronomy is moving toward the high redshift universe
  at $z >$ 6.
  Discoveries of high redshift quasars and galaxies provide prospects
 to understand the star formation and the re-ionization history of universe.
%  Furthermore, the discoveries made it possible to test galaxy formation models to the extreme
%  edge of the universe.
  The assembly history of massive galaxies at 1 $< z <$ 3 is another important topic of extragalactic study.
  It appears that massive, red galaxies are already in place at $z \sim$ 1 (\citealt{Im02}),
  and a significant build-up of massive galaxies at $z >$ 1 is hinted from the deep extragalactic surveys
 (\citealt{Drory05}; \citealt{Shim07}).
  In order to improve our understanding on the above issues, we need more work in NIR.

  Through $Y$-band observation, we can study various extragalactic objects like high redshift quasars and galaxies
 ($z >$ 6), and improve photometric
 redshift determination of red galaxies at $z > 1$ that have their 4000$\AA$ breaks redshifted to $Y$-band.
  Finding quasars and galaxies at $z > 6$ is important since they can provide ways to investigate the
 cosmic re-ionization epoch.  Going further than $z > 6.5$ is even more important, as they allow us
 to probe the universe at an earlier epoch but the number
 of known quasars at $z > 6.5$ is only handful so far (\citealt{Mortlock11}).
  Fig. \ref{fig1} demonstrates how NIR band observation is crucial especially in $Y$-band for high redshift
 quasar search. High redshift quasars and galaxies have a strong spectral break at Ly$\alpha$.
 The Ly$\alpha$ is redshifted to $z$ or $Y$-bands when the objects are at $z > 6$ or at $z > 7$, allowing
 us to use this feature to identify high redshift objects. $Y$-band enables us not only to identify the spectral break
 but also to measure the spectral shape long-ward of the Lyman break to distinguish high redshift objects from
 brown dwarfs.
  Similarly, adding an NIR point to Gamma-Ray Burst (GRB) follow-up observation in optical band is
 one of useful aspects of $Y$-band. Just like high redshift quasar search, it helps in determining
redshifts of GRBs, especially those at $z >$ 6.5, by the Lyman break dropout technique (\citealt{Steidel03}).
  Hence, recent large surveys like UKIDSS (\citealt{Lawrence07}) and Pan-STARRS surveys adopt $Y$-band.

 Considering the emergence of $Y$-band, we performed $Y$-band imaging observations
 of well-known extragalactic fields using the 1-m telescope at Mt. Lemmon Optical Astronomy Observatory (LOAO)
 in Arizona, USA, and the 1.5-m telescope at the Maidanak observatory in Uzbekistan.
  Our scientific goals are primarily three folds. The first goal is to establish the number count of $Y$-band objects at a relatively bright magnitude limit, but faint enough to go beyond the surveys such as UKIDSS Large Area Survey (LAS).
So far no work has been carried out to understand the number count study in $Y$-band. This study will be the first work
 which will establish the $Y$-band number count at bright end.
 Our number counts can be used to check the completeness of the other shallower surveys, and can offer a first look of
 the nature of sources in $Y$-band, especially at bright-end part.
 We will also check the limiting magnitudes at a given exposure time for the small telescopes we used, and gauge the usefulness of such observations for the study of distant extragalactic sources.
 The second goal is to test the selection of high redshift quasars using $i-z$ vs $z-Y$ plot.
 We obtained the $Y$-band imaging observation of the known high redshift quasars and cool dwarfs. For this study, we target $z >$ 5.8 SDSS quasars and cool dwarfs. Previous studies used $J$-band for this purpose,
 but we will show that $Y$-band can substitute the $J$-band point, therefore, making it possible to carry out
 an efficient selection of high redshift quasar candidates even with small telescopes equipped with traditional
 CCD cameras.
 The third goal is to provide photometric calibration parameters in $Y$-band.
 We will provide the basic photometric calibration
%  as well as limiting magnitude information
 in $Y$-band at the two optical observatories.

  We will first describe the facilities we used for our observation, followed by the observation, the reduction,
 and the analysis of the data. Then, the photometric calibration, and the source counts will be introduced.
 We will construct the color-color diagram for high redshift quasars and cool brown dwarfs,
 in order to directly test the usefulness of the selection method for the high redshift quasars.
 Discussion on the implications of our findings will be followed by the summary of this work.

\section{$Y$-BAND FILTER AND OBSERVATION}

 We carried out the $Y$-band imaging observations using the LOAO 1-m telescope and its 2k CCD camera, as well as the Maidanak 1.5-m telescope and its 4k CCD camera, SNUCAM (\citealt{Im10}).
These observations and the instrument characteristics are described below.

\begin{figure}[!!t]
\centering \epsfxsize=8cm
\epsfbox{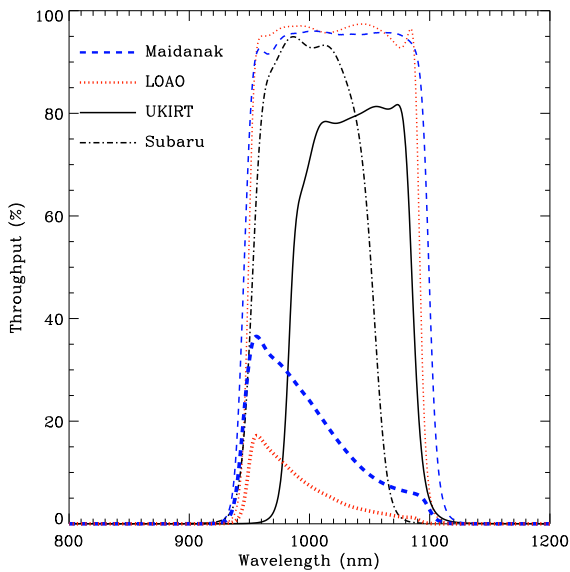}
\caption{The $Y$-band filter transmission curves and the throughputs
 of LOAO and Maidanak observatory.
 The thin lines represent the filter transmission curve only.
 The thick lines show the throughput including the detector QE
 (the blue-dashed line for SNUCAM, and the red-dotted line for
 the LOAO CCD camera).
 Note that the transmission curve of our $Y$-band filters is
 slightly different from those of the UKIRT WFCAM $Y$-band
 and z$_{R}$ filter of Subaru Suprime-Cam.
% The blue, dashed lines represent the throughput curves for SNUCAM
% and the red, dotted lines represent those of the LOAO camera.
 %The UKIRT Y-band filter transmission curve is included as a black solid line
 %for comparison. The z$_{R}$ filter of Subaru Suprime-Cam
 %has a nearly identical
% transmission curve to the UKIRT Y-band filter.
%Note that Y-band filter's wavelength region covers both UKIRT Y-band and z$_{R}$ filter of Subaru Suprime-Cam.
}
\label{fig2}
\end{figure}

\begin{table*}
%\scriptsize
\begin{center}
\centering
%\vspace{1mm}
\caption{Target fields and summary of observation}
%\label{tbl1}}
\begin{tabular}{cccccc}
%\noalign{\smallskip} \hline\hline \noalign{\smallskip}
\hline \hline \vspace{-0.3cm}\\
Target & R.A. & Dec. & Exp. time & Area & Detection limit \\
field & (J2000) & (J2000) & ($\times$ 300s) & deg$^{2}$ & 5$\sigma$ (AB mag)\\
%\noalign{\smallskip} \hline \noalign{\smallskip}
\hline
NEP\tablenotemark{c} & 02:18:00 & -07:00:00 & 18.8-21.6 & 0.1 & 21.2 \\
CFHTLS-W1\tablenotemark{c}  & 17:56:27 & 66:12:45 & 26.8-28 & 0.1 & 21.1 \\
GRB 090429B\tablenotemark{c} & 14:02:40 & 32:10:14 & 36 & 0.1 & 22.3 \\
\hline
EGS & 14:19:00 & 52:47:30 & 50 & 0.12 & 20.76 \\
FLS & 17:21:00 & 60:17:00 & 18 & 0.12 & 20.52 \\
NEP & 17:55:30 & 66:25:00 & 3-52 & 0.96 & 19.49-20.94 \\
UKIDSS & 12:00:00 & 00:00:30 & 8 & 0.12 & 19.91 \\
SDSS J113717+354956.9\tablenotemark{a} & 11:37:22 & 35:49:57 &  8 & 0.12 & 19.0 \\
SDSS J084035+562419.9\tablenotemark{a} & 08:40:35 & 56:24:20 & 18 & 0.12 & 20.5 \\
SDSS J084119+290504.4\tablenotemark{a} & 08:41:20 & 29:05:00 & 15 & 0.12 & 20.3 \\
SDSS J092721+200123.7\tablenotemark{a} & 09:27:21 & 20:01:23 & 16 & 0.12 & 20.3 \\
SDSS J125051+313021.9\tablenotemark{a} & 12:50:51 & 31:30:22 & 17 & 0.12 & 20.5 \\
SDSS J065405+652805.4\tablenotemark{b} & 06:54:05 & 65:28:05 & 14 & 0.12 & 20.3 \\
SDSS J083506+195304.3\tablenotemark{b} & 08:35:06 & 19:53:04 & 10 & 0.12 & 19.5 \\
SDSS J104335+121314.1\tablenotemark{b} & 10:43:35 & 12:13:14 & 10 & 0.12 & 19.4 \\
SDSS J121951+312849.4\tablenotemark{b} & 12:19:51 & 31:28:49 & 10 & 0.12 & 19.7 \\
SDSS J090900+652527.1\tablenotemark{b} & 09:09:00 & 65:25:27 &  8 & 0.12 & 18.5 \\
%\noalign{\smallskip} \hline \noalign{\smallskip}
\hline
\end{tabular}
\end{center}
\scriptsize
%\vspace{-0.4cm}
\begin{tabnote}
 \hskip18pt ~~~~~~~~~~~~~~~~$^a$: SDSS quasar\\
\end{tabnote}
\begin{tabnote}
 \hskip18pt ~~~~~~~~~~~~~~~~$^b$: SDSS brown dwarf\\
\end{tabnote}
\begin{tabnote}
 \hskip18pt ~~~~~~~~~~~~~~~~$^c$: Observed at Maidanak
 \end{tabnote}
\end{table*}

\subsection{$Y$-band Filters}

  The $Y$-band filters were manufactured by Asahi Spectra Inc.
 with a custom specification.
% one for the LOAO 1-m telescope,
% and another for the Maidanak 1.5-m telescope.
  The $Y$-band filter for the LOAO 1-m has a physical size of
 roughly 3 $\times$ 3 inches (square shaped),
 and the Maidanak filter has a physical size of 4 $\times$ 4 inches to accommodate the large 4k $\times$ 4k array.
 Both filters are about 1 cm thick. Two glasses are cemented each other to create the necessary transmission curve.
 The LOAO $Y$-band filter was installed on the telescope at February 2008, while the Maidanak $Y$-band filter was
 installed in the summer of 2008. After the installation of the $Y$-band filter at the Maidanak observatory,
 we also carried out test observations which will be described below.
 The filter transmission curves of our $Y$-band filters are plotted in Fig. \ref{fig2}
 which is almost identical to the Pan-STARRS $Y$-band filter. For comparison,
 $Y$-filters of UKIRT are plotted over the former.

\subsection{LOAO Observation}

\subsubsection{Targets}

 The main purposes of the LOAO observation were to understand the characteristics of the $Y$-band sky
 especially for the study of distant, extragalactic objects and checking
 the usefulness of $Y$-band observations for high-redshift quasar selection.
  Therefore, we chose target fields with the existing multi-band data.
 The fields chosen are the Extended Groth Strip (EGS; \citealt{Davis07}),
 the First Look Survey field (FLS; \citealt{Fadda06}),
 the \textit{AKARI} North Ecliptic Pole (NEP) survey field  (\citealt{Matsuhara06}; \citealt{Lee07}),
 and a part of a UKIDSS LAS field.
  The EGS is a field near the constellation Ursa Major, which boasts of a wealth of the multi-wavelength
 datasets from X-ray to radio including \textit{HST}, \textit{Chandra}, \textit{AKARI} (Song \& Im 2009),
 and a large number of spectroscopic redshifts taken by DEEP and DEEP2 surveys
 (\citealt{Vogt05}; \citealt{Im02}; \citealt{Faber07}).
 The \textit{Spitzer} FLS is the first extragalactic survey carried out by
 the \textit{Spitzer} Space Telescope at the wavelengths of 3.6, 4.5, 5.6, 8.0, 24, 70, and 160 $\mu m$,
 with the ancillary data from UV to Radio, including the Canada France Hawaii Telescopes (CFHT) MegaCam images
 in $u$-band and $g$-band taken by our group (\citealt{Shim06}).
  The UKIDSS field is chosen for comparison to our $Y$-band observation.
  The \textit{AKARI} NEP survey field is centered on the ecliptic pole,
 and it also boasts a wide range of
 multi-wavelength datasets centered on the \textit{AKARI} IR data.

 Additionally, 5 SDSS quasars at $z >$ 5.8 and 5 brown dwarfs (T and L dwarfs) were observed to verify
 the selection method to distinguish between high redshift quasars and brown dwarfs
 on a color-color diagram.
 For a few cases, $z$-band data were taken to examine the performance of
 the $z$-band imaging of the LOAO 1-m telescope.
 For standard star, we chose A0V stars from \textit{HIPPARCOS}
 (\citealt{Perryman97}) catalog because there are very few standard
 stars in $Y$-band. The A0V stars allow us to immediately calibrate
 the photometric zero-point, since A0V stars have 0 colors
 in the Vega magnitude system.
 The observed fields are summarized on the Table 1.
 The total observed area of LOAO fields spans about 2 deg$^{2}$.

\subsubsection{Observation}

  We used the LOAO 2k CCD camera system for the observations,
 which was manufactured by Finger Lakes Instrumentation with a KAF-4301E chip
 made by KODAK, Inc..  Its 24 $\mu m$ $\times$ 24 $\mu m$ scale
 translates to a pixel size of 0.64$\arcsec$,
 providing a field of view of 22$'$ $\times$ 22$'$ on the 1-m telescope.
  The telescope and the camera are remotely controlled from
 Korean Astronomy \& Space  Institute (KASI), Korea.
  The front-illuminated CCD of LOAO does not suffer from fringing
 at 1 $\mu m$ compared to back-illuminated thinned CCD chips.

   Observations were carried out on the nights of 2008 March 19-23
 and July 16-19, for a total of 9 nights. Average seeing was 2.5\arcsec,
 although some of the frames were taken with slightly out of focus.
 The data was taken with exposures exceeding 2 min.
 The moon phase was full but this did not affect the sensitivity too much
 just like any other NIR observation.
 In the period of 9 nights, the weather was clear on 5 nights,
 cloudy on 4 other.

  When guide stars were not found easily, the observation was done without guiding and the exposure per frame
 was limited to 2 minutes or less.
 The observation was done by dithering with each dither pattern
 having 20$\arcsec$ - 30$\arcsec$ steps. This procedure was necessary
 due to many hot or warm pixels
 and bad columns in the 2k CCD camera whose intensities varied with time.
 % The correction to these bad pixels required redundant observations in different pixel locations,
 %and this was achieved through the dithered observation.
  The total integration time was set to 1 hr per field (plus the overhead of about 30 min).
  In some fields, we divided the region into several tiles
 to cover a wide area.

  The standard stars were observed every night except on 16 July at various
 airmasses as the weather permitted.
  The standard star observation summary is given in Table 2.
  We also obtained the bias frames, dark frames for calibration,
 as well as the sky flats.
  Due to the low efficiency of the CCD camera at $Y$-band,
 we note that a relatively long exposure was needed for obtaining $Y$-band
 flat, which were taken usually between $R$-band and $V$-band flats.

\subsection{Maidanak Observation}

\subsubsection{Targets}

  The observed target fields are the AKARI NEP survey field,
 a small part of the XMM-LSS field of the CFHT Legacy Survey
 (\citealt{Cuillandre06}) for which $ugriz$ imaging data are available.
  The CFHTLS field overlaps with the UKIDSS Deep eXtragalactic Survey
 (DXS; \citealt{Lawrence07}) area.
  Finally, GRB 090429B was observed in $Y$-band as a part of
 GRB follow-up observation
 campaign (e.g., Lee et al. 2010) with a 3 hr total integration which
 corresponds to the deepest $Y$-band imaging data obtained by this work.
  Some of the AKARI NEP fields were covered at Maidanak as well.
  The observed fields are summarized in Table 1.

\subsubsection{Observation}

 The observation was carried out using SNUCAM on the 1.5-m telescope
 (Im et al. 2010) during August 30, 2008
 through September 4 for 6 nights, except for the GRB 090429B field.
 SNUCAM is a cryo-cooled CCD camera with a back-illuminated,
 thinned 4k $\times$ 4k CCD chip manufactured by Fairchild, Inc.
 On the 1.5-m telescope, SNUCAM has a pixel size of 0.266$\arcsec$,
 and the field of view of 18$'$ $\times$ 18$'$.
 The weather conditions during the observation were generally good,
 but we accumulated the data
 worth about two nights due to operation problems with the telescope.
 The average seeing was 1.2$\arcsec$, although some of the frames were taken
 slightly out of focus.
 The exposure per frame was limited to 2 minutes or less because of tracking
 instability and the need for dithering.
 The observation was done by dithering with each dither pattern having
 20$\arcsec$ - 30$\arcsec$ steps.
 This procedure was necessary due to fringing which is typical for
 a back-illuminated, thinned CCD chip.
 We observed A0V stars from the \textit{HIPPARCOS} catalog as standard stars
 for most cases and a UKIRT faint standard star (P272-D) when we observed GRB 090429B.
 We also obtained the bias frames, and flat frames for calibration.

\begin{table*}
%\scriptsize
\begin{center}
\centering
%\vspace{1mm}
\caption{Observed standard star list}
%\label{tbl1}}
\begin{tabular}{cccccccccc}
%\noalign{\smallskip} \hline\hline \noalign{\smallskip}
\hline \hline \vspace{-0.3cm}\\
Date & Standard & R.A. & Dec. & Exp. time & Airmass & Zero & $V$ mag & E($B-V$) & A$_{V}$ \\
(2008) & star & (J2000) & (J2000) & [second] & sec(Z) & point &
\textit{Vega}) & [mag] & [mag] \\
%\noalign{\smallskip} \hline \noalign{\smallskip}
\hline
Mar. 20 & HIP 32549 & 06:47:28 & 44:19:53 & 20 & 1.256 & 18.10 & 8.24 & 0.066 & 0.219 \\
       & HIP 71172 & 14:33:23 & 69:37:44 & 60 & 1.075 & 18.41 & 6.06 & 0.016 & 0.053 \\
Mar. 21 & HIP 32549 & 06:47:28 & 44:19:53 & 20 & 1.258 & 18.25 & 8.24 & 0.066 & 0.219 \\
       & HIP 33297 & 06:55:34 & 08:19:27 & 10 & 1.096 & 18.14 & 9.74 & 0.220 & 0.731 \\
       & HIP 42853 & 08:43:56 & 19:02:03 & 20 & 1.071 & 18.53 & 8.30 & 0.029 & 0.095 \\
       & HIP 71172 & 14:33:23 & 69:37:44 & 20 & 1.193 & 18.34 & 6.06 & 0.016 & 0.053 \\
Mar. 22 & HIP 32549 & 06:47:28 & 44:19:53 & 20 & 1.270 & 18.20 & 8.24 & 0.066 & 0.219 \\
       & HIP 34107 & 07:24:20 & 01:29:18 & 10 & 1.182 & 18.28 & 6.57 & 0.364 & 1.206 \\
       & HIP 42853 & 08:43:56 & 19:02:03 & 20 & 1.040 & 18.62 & 8.30 & 0.029 & 0.095 \\
       & HIP 42853 & 08:43:56 & 19:02:03 & 20 & 1.777 & 18.57 & 8.30 & 0.029 & 0.095 \\
       & HIP 71172 & 14:33:23 & 69:37:44 & 20 & 1.221 & 18.40 & 6.06 & 0.016 & 0.053 \\
Mar. 23 & HIP 32549 & 06:47:28 & 44:19:53 & 20 & 1.261 & 18.21 & 8.24 & 0.066 & 0.219 \\
       & HIP 34107 & 07:24:20 & 01:29:18 & 12 & 1.168 & 18.23 & 6.57 & 0.364 & 1.206 \\
       & HIP 42853 & 08:43:56 & 19:02:03 & 20 & 1.062 & 18.55 & 8.30 & 0.029 & 0.095 \\
       & HIP 42853 & 08:43:56 & 19:02:03 & 25 & 1.378 & 18.54 & 8.30 & 0.029 & 0.095 \\
Mar. 24 & HIP 32549 & 06:47:28 & 44:19:53 & 20 & 1.272 & 18.23 & 8.24 & 0.066 & 0.219 \\
       & HIP 34107 & 07:24:20 & 01:29:18 & 12 & 1.182 & 18.24 & 6.57 & 0.364 & 1.206 \\
       & HIP 42853 & 08:43:56 & 19:02:03 & 30 & 1.042 & 18.53 & 8.30 & 0.029 & 0.095 \\
       & HIP 42853 & 08:43:56 & 19:02:03 & 30 & 1.596 & 18.48 & 8.30 & 0.029 & 0.095 \\
Jun. 18 & HIP 88429 & 18:03:14 & 19:36:47 & 10 & 1.360 & 18.00 & 6.41 & 0.099 & 0.327 \\
Jun. 19 & HIP 65599 & 13:26:58 & 11:54:30 & 20 & 1.081 & 18.01 & 7.99 & 0.027 & 0.091 \\
       & HIP 75230 & 15:22:23 & 12:34:03 & 12 & 1.107 & 18.04 & 6.41 & 0.043 & 0.144 \\
       & HIP 88429 & 18:03:14 & 19:36:47 & 10 & 1.176 & 18.12 & 6.41 & 0.099 & 0.327 \\
       & HIP 88429 & 18:03:14 & 19:36:47 & 10 & 1.557 & 18.07 & 6.41 & 0.099 & 0.327 \\
Jun. 20 & HIP 65599 & 13:26:58 & 11:54:30 & 20 & 1.077 & 18.08 & 7.99 & 0.027 & 0.091 \\
       & HIP 88429 & 18:03:14 & 19:36:47 & 10 & 1.373 & 18.06 & 6.41 & 0.099 & 0.327 \\
       & HIP 88429 & 18:03:14 & 19:36:47 & 10 & 1.636 & 18.11 & 6.41 & 0.099 & 0.327 \\
\hline \\
Aug. 30 & HIP 98460\tablenotemark{a} & 20:02:05 & 12:19:19 & 10 & 1.378 & 19.10 & 9.38 & 0.203 & 0.674 \\
       & HIP 98460\tablenotemark{a} & 20:02:05 & 12:19:19 & 20 & 1.208 & 19.17 & 9.38 & 0.203 & 0.674 \\
       & HIP 111538\tablenotemark{a} & 22:35:48 & 40:05:27 & 20 & 1.149 & 18.94 & 9.48 & 0.127 & 0.570 \\
       & HIP 111538\tablenotemark{a} & 22:35:48 & 40:05:27 & 20 & 1.056 & 18.94 & 9.48 & 0.127 & 0.570 \\
       & HIP 582\tablenotemark{a}   & 00:07:05 & 60:23:33 & 20 & 1.266 & 19.24 & 9.52 & 0.959 & 3.178 \\
Aug. 31 & HIP 98460\tablenotemark{a} & 20:02:05 & 12:19:19  & 20 & 1.462 & 19.10 & 9.38 & 0.203 & 0.674 \\
       & HIP 98460\tablenotemark{a} & 20:02:05 & 12:19:19  & 20 & 1.197 & 19.10 & 9.38 & 0.203 & 0.674 \\
       & HIP 111538\tablenotemark{a} & 22:35:48 & 40:05:27 & 20 & 1.207 & 18.94 & 9.48 & 0.127 & 0.570 \\
       & HIP 111538\tablenotemark{a} & 22:35:48 & 40:05:27 & 20 & 1.051 & 18.93 & 9.48 & 0.127 & 0.570 \\
       & HIP 582\tablenotemark{a}    & 00:07:05 & 60:23:33 & 20 & 1.271 & 19.26 & 9.52 & 0.959 & 3.178 \\
Apr. 29 & P272-D\tablenotemark{a} & 14:58:33.1 & 37:08:33 & 30 & 1.050 & 20.12 & 11.839 & 0.020 & 0.067 \\
(2009) &   (MKO)             &            &          &    &       &       & ($Y$ mag)&     &       \\
\hline
\end{tabular}
\end{center}
\scriptsize
$^a$: Observed at Maidanak
\end{table*}

\begin{figure}[!t]
\centering \epsfxsize=8cm
\epsfbox{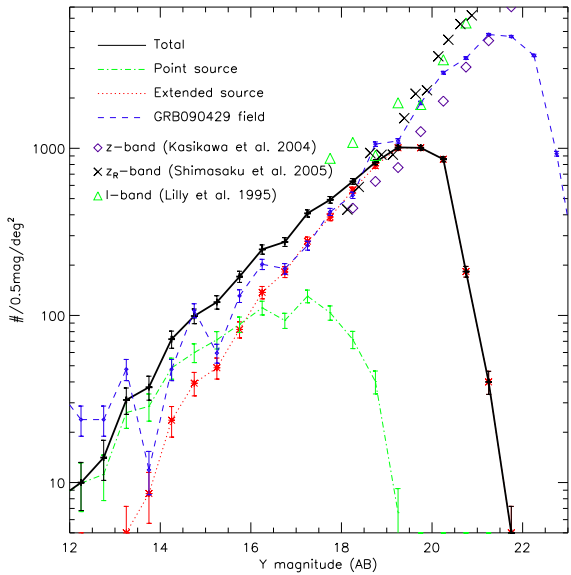}
\caption{The number counts of $Y$-band sources.
 For comparison, the observed number counts from literatures are shown
 for $I$-band (green triangles), $z$-band (black crosses)
 and $z_{R}$-band (purple diamonds). The shallow $Y$-band counts
 ($Y < 18.5$ mag) are constructed from all the fields we observed
 (total area of $\sim$2 deg$^{2}$). The number count in
 the GRB 090429B field, which is the deepest among the observed fields,
 is shown with the blue dashed line (0.08 deg$^{2}$).}
\label{fig3}
\end{figure}

\section{DATA REDUCTION}

\subsection{Image Reduction and Stacking}
We used IRAF packages for pre-processing and stacking of the data.
We describe the data reduction procedure below.

\subsubsection{LOAO data}
 \begin{figure}[!!!t]
\centering \epsfxsize=8cm
\epsfbox{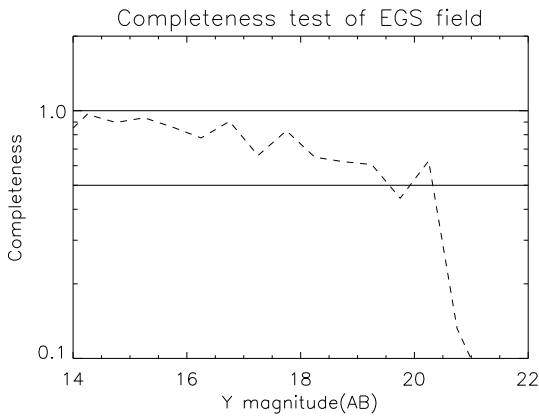}
\caption{Completeness test of an EGS field}
\label{fig4}
\end{figure}

% We reduced the LOAO data using IRAF packages.
 The reduction follows a standard procedure of the bias and the dark subtraction, followed by
 the flat-fielding. To create a master bias image for each night, 40 bias frames were combined.
 Master darks are constructed from the darks taken with the same exposure times as the science frames.
 The dark subtraction was crucial to remove many of the bad pixels
 and bad columns.
 Flat fields were constructed using skyflats taken during twilight.
 The peak-to-peak variation of the skyflat is measured to be 0.8 - 1.4.

 The reduction process removes most of the artificial features on the image, but some of the
 time-variable bad pixels remain which are removed later by stacking of the dithered frames.
 A large scale pattern remains at 3\% level, peak-to-peak.
 We suspect that this large scale pattern arises mostly from imperfect flat-fielding
 where the flat-field data rely on the data taken in a couple of nights.
 Overall, we believe that our photometry has an intrinsic uncertainty of about 3\% level (0.03mag) due to the flat-field variation.

  For registration of the reduced, dithered images,
 we selected non-saturated stellar objects
 with good Gaussian profiles as registration stars,
 and performed the registration and image shift
 using  \texttt{imalign} procedure.
  The registered images were stacked in median after subtracting the background level determined
 to be a mode of the pixel value distribution of each image.
 Cosmic ray rejection (\texttt{crrejct}) was applied during the stacking process.

% then checked its physical coordinate using other images and catalogs such as DSS(Digitized Sky Survey) images and the USNO-B1(United State Naval Observatory; \citealt{Monet03}).

\subsubsection{Maidanak data}

 The procedure of the Maidanak SNUCAM data reduction follows that of \citet{Jeon10}.
 This is similar to the LOAO data reduction procedure, but with a few differences.

 The SNUCAM data do not require dark subtraction since SNUCAM has virtually no dark
 current at the operating temperature. On the other hand, it requires additional subtraction of
 the background level in the four quadrant readout channels. Also, the fringing pattern
 needs to be removed which is severe in $Y$-band images.

 Flat fields were constructed using skyflats taken during twilight. The number of skyflats was typically 3 - 5,
 and these were combined by setting the \texttt{flatcombine} parameter of zero to ``mode".
 The peak-to-peak variation of the skyflat is measured to be 0.98 - 1.02.
 The reduction process removes most of the artificial features on the image, although a large scale pattern
 remains at more than 5\% level due to fringing.
 To correct the fringe patterns, we created a master fringe frame by stacking all of the bias-subtracted,
 flat-fielded object frames in median without performing registration. The master fringe frame was scaled
 by the sky-background level of each science frame and subtracted from the object frame (see also \citealt{Jeon10}). After the fringe correction, the images were registered and stacked in the same way as the LOAO images.

\subsection{Photometry Calibration}

 We measured the flux of the standard stars to get photometric parameters in $Y$-band.
 A standard equation as below is used for deriving the photometry calibration parameters, $\xi$ (zero-point), $\kappa$ (atmospheric extinction coefficient).

\begin{equation}
 \xi = m_{0} + 2.5\,log(DN/sec) + \kappa (X-1)
\end{equation}

 Here $m_{0}$ is a known V magnitude of an A0V star, and
 $X$ is the airmass which is equal to sec(z) where z is the zenith distance.
 By definition, A0V star's color is zero. The color-term is ignored here.
 Aperture photometry of a standard star observed twice in a day was used to
 calculate $\kappa$ and $\xi$ at different airmass values.
% For example, a $\kappa$ was derived for a certain photometric standard
% when such a star was observed at two airmass values.
 The total flux of a star is obtained with an aperture diameter of
 14$\arcsec$, to minimize the flux loss.
 The sky was estimated using an annulus of 28$\arcsec$ diameter, with a width of 7$\arcsec$. The photometry calibration
 results are summarized in Table 2. We note that the atmospheric extinction coefficient, $\kappa$, has
 a value between  0.05 and 0.1, and we take the average of these values as a measure of $\kappa$ for
 the nights where  no standard star data were taken. In this way, we find that the atmospheric extinction coefficients
 are $\kappa$ = 0.087 for LOAO, $\kappa$ = 0.1 for Maidanak observatory. The zero point, $\xi$, is calculated for
 each standard star. We find that the zero points vary between different standard stars by as much as 0.5 mag
 in some cases.
 We attributed this to be due to the weather conditions, the uncertainty in E(B-V) corrections,
 as well as the variability of standard stars in some cases. For example, the data were taken with thin clouds
 present for the March 18-20 data, which resulted in a relatively bright zero point value.
 We also found that some of the standard star data were a little out-of-focus, and the photometry of such stars
 seem spurious. Overall, the zero-points were allowed to vary between day-to-day,
 and it ranges between $\xi$ = 18.1 - 18.5 in LOAO, $\xi$ = 18.9 - 19.3 in Maidanak.
 We expect the uncertainty in the zero-point to be within 0.1 magnitude through comparison of zero-points
 from various standard stars within a single night.

\subsection{Astrometry Calibration}

 Astrometry was done with SCAMP and SWarp software
 (Terapix; \citealt{Bertin08})
 with stars in the United State Naval Observatory (USNO-B1; \citealt{Monet03})
 catalog as reference stars.
 After astrometry calibration, the rms error of astrometry is computed to be
 less than 0.35$\arcsec$ from the SCAMP software.
%  Here, we set search radius 2$\arcsec$ for all the cross identification.
  After the astrometry calibration we updated the header of image files
 using SWarp.

 \begin{figure}[!t]
\centering \epsfxsize=8cm
\epsfbox{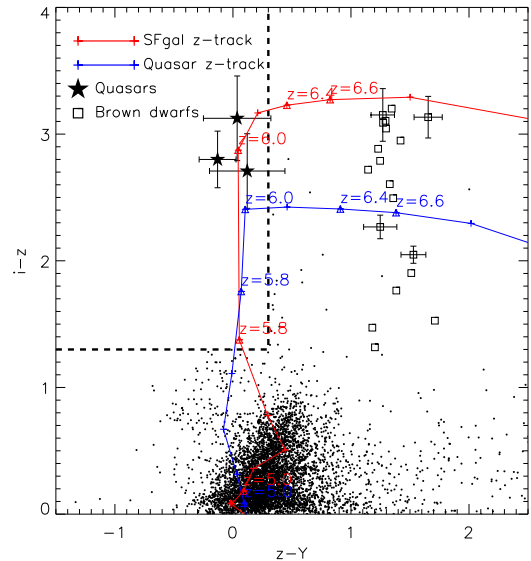}
\caption{Color-color plot in $i-Y$ vs $z-Y$. Black stars are known quasar at $z$ $\sim$ 6. Open squares with error bars are brown dwarfs and other open squares without error bars are L/T dwarf stars taken from Hewett et al. (2006). Black dots are other sources from the observed data such as stars and galaxies.
 Blue solid line is an evolutionary track of star forming galaxy and red line is that of a quasar.}
\label{fig5}
\end{figure}

\subsection{Source Detection and Extraction}

 SExtractor (\citealt{Bertin96}) was used for the object detection and
 photometry.
 The magnitude zero-points of standard stars are adopted as in Table 2 for each night. The DETECTION THRESHOLD is
 set to 1.2$\sigma$ and the MINIMUM AREA to be 4 pixels.
 The background mesh size of 32 $\times$ 32 pixels is adopted for
 estimating local background.
 The parameters were chosen so that we minimize the spurious detections
 as well as making sure any obvious objects were not missed through
 the eye-ball inspection of the image.
 In general, about 2000 objects are detected per field with $\sim$ 1 hr total integration.
  We adopt the Auto magnitude from SExtractor. To transform the Vega magnitude
 system to AB magnitude system,  we adopted
 the relation of $Y_{AB}$ mag= $Y_{Vega}$ mag + 0.63 from \citet{Hewett06}.
%  We also matched our $Y$-band data with existing catalogs from various sources containing optical photometry and redshifts.
%  For such a purpose, we use the SDSS DR7 imaging catalog (\citealt{Abazajian08}),
% CFHT-NEP matched catalog (\citealt{Hwang07}) and UKIDSS DR2 PLUS (\citealt{Warren07}) catalog, and DEEP2 redshift catalog. Matching radius was chosen to be 1.5$\arcsec$. Note that hereafter, all magnitude system is AB mag system unless otherwise mentioned.

\section{RESULTS}

\begin{table*}
\scriptsize
\begin{center}
\centering
%\vspace{-3mm}
\caption{Number counts}
%\label{tbl1}}
\begin{tabular}{rrrrr}
%\noalign{\smallskip} \hline\hline \noalign{\smallskip}

\hline \hline \\
$Y$ magnitude & Total \tablenotemark{a} & Extended source \tablenotemark{a} & Point source \tablenotemark{a} & GRB 090429B \tablenotemark{a} \\
(AB mag) & (deg$^{-2}$) & (deg$^{-2}$) & (deg$^{-2}$) & field (deg$^{-2}$)\\
%\noalign{\smallskip} \hline \noalign{\smallskip}
\hline
11.5$-$12.0 &  8.0 $\pm$ 2.8    & 0.0 $\pm$ 0.0      &  8.0 $\pm$ 2.8 & 35.7 $\pm$ 5.9\\
12.0$-$12.5 & 10.0 $\pm$ 3.2    & 0.1 $\pm$ 0.3      &  9.9 $\pm$ 3.2 & 23.8 $\pm$ 4.8\\
12.5$-$13.0 & 14.1 $\pm$ 3.8    & 2.9 $\pm$ 1.7      & 11.2 $\pm$ 3.4 & 23.8 $\pm$ 4.8\\
13.0$-$13.5 & 31.2 $\pm$ 5.6    & 5.0 $\pm$ 2.2      & 26.2 $\pm$ 5.1 & 47.6 $\pm$ 6.9\\
13.5$-$14.0 & 37.2 $\pm$ 6.1    & 8.6 $\pm$ 2.9      & 28.6 $\pm$ 5.3 & 11.9 $\pm$ 3.45\\
14.0$-$14.5 & 72.1 $\pm$ 8.5    & 23.6 $\pm$ 4.9     & 48.5 $\pm$ 7.0  & 47.6 $\pm$ 6.9\\
14.5$-$15.0 & 99.2 $\pm$ 10.0   & 39.3 $\pm$ 6.3     & 59.9 $\pm$ 7.7 & 107.1 $\pm$ 10.35\\
15.0$-$15.5 & 120.1 $\pm$ 11.0  & 48.6 $\pm$ 7.0     & 71.5 $\pm$ 8.5 & 59.5 $\pm$ 7.7\\
15.5$-$16.0 & 170.7 $\pm$ 13.1  & 82.4 $\pm$ 9.1     & 88.4 $\pm$ 9.4 & 130.9 $\pm$ 11.4\\
16.0$-$16.5 & 248.4 $\pm$ 15.8  & 137.3 $\pm$ 11.7   & 111.1 $\pm$ 10.5 & 202.4 $\pm$ 14.2\\
16.5$-$17.0 & 275.3 $\pm$ 16.6  & 182 $\pm$ 13.5     & 93.3 $\pm$ 9.7 & 190.5 $\pm$ 13.8\\
17.0$-$17.5 & 408.2 $\pm$ 20.2  & 277.8 $\pm$ 16.7   & 130.4 $\pm$ 11.4 & 261.9 $\pm$ 16.2\\
17.5$-$18.0 & 492.1 $\pm$ 22.2  & 388.2 $\pm$ 19.7   & 103.8 $\pm$ 10.2 & 416.7 $\pm$ 20.4\\
18.0$-$18.5 & 633.9 $\pm$ 25.2  & 562.1 $\pm$ 23.7   & 71.8 $\pm$ 8.5 & 523.8 $\pm$ 22.9\\
18.5$-$19.0 & 825.9 $\pm$ 28.7  & 785.7 $\pm$ 28.0   & 40.2 $\pm$ 6.3 & 1059.5 $\pm$ 32.5\\
19.0$-$19.5 & 1012.4 $\pm$ 31.8 & 1005.8 $\pm$ 31.7  & 6.6 $\pm$ 2.6  & 1107.1 $\pm$ 33.3\\
19.5$-$20.0 & 1006.6 $\pm$ 31.7 & 1006.6 $\pm$ 31.7  &                & 1869.1 $\pm$ 43.2\\
20.0$-$20.5 & 860.0 $\pm$ 29.3  & 860 $\pm$ 29.3     &                & 2833.3 $\pm$ 53.2\\
20.5$-$21.0 & 183.0 $\pm$ 13.5  & 183 $\pm$ 13.5     &                & 3464.3 $\pm$ 58.8\\
21.0$-$21.5 & 40.0 $\pm$ 6.3    & 40.0 $\pm$ 6.3     &                & 4797.6 $\pm$ 69.2\\
21.5$-$22.0 & 5.0 $\pm$ 2.2     & 5.0 $\pm$ 2.2      &                & 4666.7 $\pm$ 68.3\\
22.0$-$22.5 &                   &                    &                & 3595.2 $\pm$ 59.9\\
22.5$-$23.0 &                   &                    &                & 928.6 $\pm$ 30.4\\
\hline
\end{tabular}
\end{center}
\scriptsize
\begin{tabnote}
 \hskip18pt ~~~~~~~~~~~~~~~~~~~~~~~~~~~$^a$ The errors are calculated assuming Poisson noise.  \\
\end{tabnote}
  \end{table*}

\subsection{Number Counts}

 We derived the $Y$-band number counts from various fields (Fig.  \ref{fig3} and Table 3). Note that the error bars represent the Poisson errors.
% We also combine the number counts from all these fields by combining them all together.
 When deriving the number counts, we used counts from each field
 down to 10$\sigma$ detection limit. The number counts are also divided into
 the point and the extended sources.
  Point sources are classified by STELLARITY value larger than 0.8 given
 from SExtractor (down to $Y \sim 19$ mag), the remainder which have value less than 0.8 are classified
 as extended sources.
 We plot the number counts of the point sources with a green dot-dashed line
 and the extended sources with a red dotted line. The number counts of the
 GRB 090429B field was plotted independently (blue dashed line), since this field is much
 deeper than the other fields.
 We summarize the number counts in Table 3.
 The figure shows that the number counts at the bright end is dominated by stars and at the faint end is dominate by
 extended sources (galaxies). The faint-end number counts somewhat mimic
 those of $I$-band number counts.
 The number of sources down to $Y$=20 mag is roughly 800 per deg$^{2}$.

\subsection{Completeness Test}

 We test the completeness using an EGS field.
 We created artificial stars and galaxies at magnitude range of 12 - 22 and
 placed them on the observed image, using
 the IRAF tasks of \texttt{starlist}, \texttt{gallist} and \texttt{mkobjects}.
 40\% of the extended sources are chosen as elliptical galaxies and the others
 as disk galaxies.
 Then, we checked to see how many of them could
 be detected with the same configuration of source detection as what
 we adopted for the source detection parameters as described in Section
 3.4.
 Fig. \ref{fig4} shows that this field has 50\% completeness at 20.5 mag
 which is consistent with 5$\sigma$ depth of the EGS field
 in Table 1.

\subsection{High Redshift Quasar Selection}

 One of the main motivations for this $Y$-band study is to test the effectiveness of
 the $Y$-band imaging for the selection of high redshift quasars.
 Here, we plot the $i-z$  versus $z-Y$ colors of sources in the extragalactic fields in Fig. \ref{fig5}.
 In the plot, we show  high redshift quasars (z $\sim$ 5.8) with filled stars,
 and T/L dwarfs with squares. The expectation from Fig. \ref{fig1} is
 that high redshift quasars have very
 red $i-z$ colors, while their $z-Y$ colors should be roughly flat (around 0).
 On the other hand, the steep SED slope of T/L dwarfs make them look very red
 in $i-z$ color like high redshift quasars,
  but they are also red in $z-Y$ colors.
  This point is clearly demonstrated in Fig. \ref{fig5}. Both high redshift
 quasars and T/L dwarfs are
 very red objects in $i-z$ colors, but in the plot we can separate them
from each other
 since the cool dwarfs are redder than the high redshift quasars in $z-Y$
 colors.

 %  One strange object is far from zero color value open pentagram symbol is a T dwarf. We found its image is out of focus, so to be extended, and we are doubtful about the validity of its photometry. Nevertheless, we include this object in the plot for the sake of completeness.
 Previously, Fan et al. (2001, 2002) established a color-color space
 to select high redshift quasar candidates
 using the $i-z$ versus $z-J$ color-color diagram. Our result here
 demonstrates that a similarly effective exclusion
 of T/L dwarf stars from high redshift quasar candidates is possible,
 even with a $i-z$ versus $z-Y$ color-color diagram.

 Our result justifies the use of a moderate sized telescope
 with a simple CCD camera, even for the selection of the highest redshift quasar candidates.
%  These color-color plots also include points which satisfy the T/L dwarf selection criteria as well as
%the high redshift quasar selection.
% These objects were examined further. We selected total 31 objects which satisfied the criteria of $i-z$ $>$ 2
% and $z-Y$ $<$ 0 from color-color diagram. The inspection of the image shows some of the objects are
% stellar objects like shown as orange point on the plot.
% The other 13 objects could be recognized with eyes, typical $Y$ magnitude of them is 21 - 22 AB mag with somewhat large magnitude error bigger than 0.3, so after exclusion of objects which have bigger magnitude error than 0.3, there is no candidate for high-z quasar from our observation.
% It is known that we can find 1 or less high-z quasar (z $>$ 6) per 1 deg$^{2}$ area down to 23 AB magnitude. Our total observation area is 2.5 deg$^{2}$, but it lacks sufficient depth. We need deeper and wider $Y$-band imaging to find high-z quasars.

\section{DISCUSSION}

\subsection{Detection Limits of LOAO and Maidanak Telescopes}

 Small telescopes are on disadvantage when studying faint sources at high redshift
 compared to large telescopes.
 However, for sources sufficiently bright such as quasars and GRBs,
 small telescopes can make an impact as long as the integration time to provide enough S/N is
 not too long.
 This is because, the observing times are much more readily available for small telescopes.

\begin{figure}[!t]
\centering \epsfxsize=8cm
\epsfbox{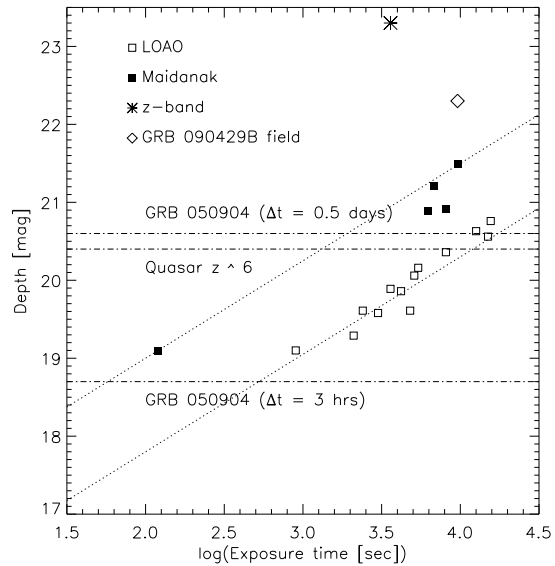}
\caption{5$\sigma$ detection limits (AB mag) at a given exposure time.
 Open squares are for the LOAO limits, asterisks are for the Maidanak limits.
 A quasar at z $\sim$ 6 and GRB 050904 at z $\sim$ 6.3 are marked as horizontal
 dot-dashed lines.
 Diagonal dotted lines represent detection limits for Maidanak (the upper line)
 and for LOAO (the lower line). We indicate the $z$-band depth of 2 hr
 integration time at Maidanak for comparison which is located above
 23 magnitude. The deepest $Y$-band Maidanak detection limit
 indicates the value derived from GRB 090429B.}
\label{fig6}
\end{figure}

 Fig. \ref{fig6} shows the integration time versus 5$\sigma$ detection limits (in AB magnitude)
 for the LOAO and the Maidanak facilities. We see that the Maidanak telescope goes deeper,
 but not much deeper than the LOAO depth, despite the larger aperture of the mirror. The main culprit
 for the inefficiency of the Maidanak is the dusty primary mirror which is
 sometimes way over-due for the mirror coating and cleaning.
 The mirror of Maidanak 1.5m telescope was cleaned in 2009,
  after that we can see a marked improvement in depth through the observation
 of the GRB 090429B field.

  The improvement is 1 mag deeper than before clearing in $Y$-band.

% Nevertheless, the excellent seeing conditions at Maidanak,
% the highly QE of SNUCAM, and the readily available observing resources can potentially make
% the Maidanak observatory a prime observing site for the future $Y$-band imaging. In principle,
% one can reach $Y$=22.5 mag using a whole night of observation ($\sim$8 hrs)
% at the seeing condition of $\sim$ 1$\arcsec$.

  High redshift quasar candidates, such as SDSS $i$-band dropouts at 5.8 $< z <$ 6.5, typically have $z <$ 20.2 AB mag.
  Therefore, it is sufficient to detect objects brighter than $Y \sim$ 20 AB mag in order to enable efficient filtering of L/T-dwarfs.
  Actually, the magnitude limit can be brighter for the filtering of the L/T-dwarfs,
  since they are much brighter in the $Y$-band than z $\sim$ 6 quasars for a given $z$-band magnitude.
  The required $Y$-band magnitude limit can be achieved at about 1 hr exposure
 time on a clear night with the LOAO 1-m telescope even at a bad seeing,
 and much less than 1 hr with SNUCAM at the 1.5m assuming
 that it has a clean mirror.
  We expect that the observations at LOAO and Maidanak using $Y$-band filters
 can greatly help reduce our reliance on the $J$-band imaging for high redshift quasar candidate selection.

  GRBs at high redshift can be detected with $Y$-band too. For example, GRB 050904 at $z=6.3$ (Haislip et al. 2006)
 was as bright as $Y \sim 18.7$ AB mag at 3 hrs post-burst, and $Y \sim 20.6$ AB mag at 0.5 days post-burst. Detection
 of such an event is possible at a few minutes (Maidanak) to a few tens of minutes (LOAO) shortly after the burst,
 and at 1$-$3 hrs exposure time even at $\sim 0.5$ days post-burst.

\subsection{Promises of $Y$-band Imaging with Small Telescopes}

 There are several interesting sciences that can be potentially carried out
 by a moderate sized telescope.
 First, a rapid response observation of GRB afterglow events,
 especially for those suspected to be occurring at high redshift,
 is interesting. The GRB afterglow has been identified out to z $\sim$ 9.4
 (Cucchiara et al. 2011), and it is in principle possible to identify an afterglow
 at much higher redshifts.
  For GRB afterglows such as $i$-band or $z$-band dropouts,
 a combination of $i$-band / $z$-band / $Y$-band observation
 should be able to identify their redshift nature.
  Even at a moderate $\sim$10 minute exposure time, our observations suggest
 that we can detect  GRB afterglows at 18-19 magnitude level using
 the LOAO 1.0-m or  Maidanak 1.5-m telescopes.

  $Y$-band observation can also provide an interesting data point in studying
 the SED of dusty GRBs. Recently, a high redshift GRB at $z \sim 5$ was
 identified to harbor supernova-produced dust (\citealt{Perley10}; \citealt{Jang11}).
 $Y$-band was one of key data points for distinguishing the extinction curve of
 supernova-produced dust from the dust extinction curves of
 other origins.

 %Next interesting project would be the planet transit. The habitable zone of the M-stars is very close to the host-star.
% The sizes of the M-stars are small, therefore, the transit signal from the earth-size planets on the M-stars
 %can be relatively easier detect than other cases.
 %$Y$-band imaging on a 1-m or 1.5-m telescope has several advantages in observing transit events.
 %One of the problems with the transit observations of nearby stars is that the stars are too bright
 %for efficient observations. The reduced sensitivity at $Y$-band is actually advantageous in this sense to avoid
 %the saturation. Also, M-stars are brighter in the $Y$-band (or longer wavelengths),
 %so that the $Y$-band observation is best-suited for such effort. The limb-darkening effect is also minimal
 %at the longer wavelength and the $Y$-band offers an advantage in that respect too.

\section{SUMMARY}

  We carried out $Y$-band imaging observations of extragalactic fields using 1-m class telescopes at
 LOAO and Maidanak observatory. The total observed area was $\sim$ 2 deg$^{2}$,
 and we reached $Y$ = 23 AB mag depth with 3 - 4 hour integration with
 the Maidanak 1.5-m telescope.
 We obtained photometric calibration data and derive the atmospheric
 extinction coefficient $\kappa$ = 0.087 at LOAO, and
 $\kappa$ = 0.1 at Maidanak observatory in $Y$-band from A0V standard stars.
  We present the number counts of $Y$-band sources which provide a useful
 constraint at the bright end of the number count.
  At a very bright end, point-like sources dominate the number count,
 while the faint end number count is dominated by galaxies.
  The color-color diagram $i-z$ vs. $z-Y$ is confirmed to be
 effective for discriminating high redshift quasars (z $>$ 6) from brown dwarfs.
 With the versatility and the availability of 1-m class telescopes, an 1-m class telescope can be
 useful for studying high redshift
 quasars and GRBs, and other red sources, once its capability is enhanced with
 a unique device such as $Y$-band filter.

\acknowledgments
 We acknowledge the support from the National Research Foundation of Korea (NRF) grant funded by the
 Korean government (MEST), No. 2010-0000712.
 This work made use of the data taken with the
 LOAO 1-m telescope operated by KASI and the Maidanak 1.5-m telescope
 in Uzbekistan.
 We thank CEOU members for their help with useful advices, and the staffs
 of the Maidanak observatory and KASI for their assistance during
 our observation.
%-----------------------------------------------------------------------

%-----------------------------------------------------------------------
%-----------------------------------------------------------------------
%-------------------------------------------------------------------

\begin{thebibliography}{}

%------------------------------------------------------------------------
\bibitem[Abazajian et al.(2009)]{Abazajian08}
Abazajian., et al. 2009, The Seventh Data Release of the Sloan Digital Sky Survey, ApJS, 182, 543

\bibitem[Bertin \& Arnouts(1996)]{Bertin96}
Bertin, E., \& Arnouts. 1996, SExtractor: Software for Source Extraction, A\&AS, 117, 93

%\bibitem[Bertin (2006)]{Bertin06}
%Bertin, E., 2006, Automatic Astrometric and Photometric Calibration with SCAMP, ASPC, 351. 112B

\bibitem[Bertin(2008)]{Bertin08}
Bertin, E. 2008, SWarp v2.17.0 Users guide

\bibitem[Burrows et al.(2003)]{Burrows03}
Burrows, A., et al. 2003, Beyond the T Dwarfs: Theoretical Spectra, Colors, and Detectability of the Coolest Brown Dwarfs, ApJ, 596, 587

%\bibitem[Chambers \etal (2006)]{Chambers06}
%Chambers, \& Kenneth C., 2009,
%Pan-STARRS Telescope \#1 Status and Science Mission, AAS, 21330107C

\bibitem[Chiu et al.(2007)]{Chiu07}
Chiu, K., et al. 2007, The optical and Near-Infrared Properties of 2837 Quasars in the United Kingdom Infrared Telescope Infrared Deep Sky Survey, MNRAS, 375, 1180

\bibitem[Colless(1995)]{Colless95}
Colless, M., et al. 2001, The 2dF Galaxy Redshift Survey: spectra and redshifts, MNRAS, 328, 1039


\bibitem[ Cucchiara et al.(2011)]{Cucchiara11}
 Cucchiara, A., et al. 2011, A Photometric Redshift of z $\sim$ 9.4 for GRB 090429B, ApJ, 736, 7

\bibitem[Cuillandre \& Bertin(2006)]{Cuillandre06}
Cuillandre, J.-C., \& Bertin, E. 2006, CFHT Legacy Survey (CFHTLS) : A Rich Data Set, SF2A, CONF, 265

\bibitem[Davis et al.(2007)]{Davis07}
Davis, M., et al. 2007, The All-Wavelength Extended Groth Strip International Survey (AEGIS) Data Sets, ApJ, 660, L1

\bibitem[Drory et al.(2005)]{Drory05}
Drory, N., et al. 2005, The Stellar Mass Function of Galaxies to z $\sim$ 5 in the FORS Deep and GOODS-South Fields, ApJ, 619, L131

\bibitem[Faber et al.(2007)]{Faber07}
Faber, S. M., et al. 2007, Galaxy Luminosity Functions to z $\sim$ 1 from DEEP2 and COMBO-17: Implications for Red Galaxy Formation, ApJ, 665, 265

\bibitem[Fadda et al.(2006)]{Fadda06}
Fadda, D., et al. 2006, The \textit{Spitzer} Space Telescope Extragalactic First Look Survey: 24 $\mu m$ Data Reduction, Catalog, and Source Identification, AJ, 131, 2859

\bibitem[Fan et al.(2000)]{Fan00}
Fan, X., et al. 2000, The Discovery of a Luminous Z=5.80 Quasar from the Sloan Digital Sky Survey, AJ, 120, 1167

\bibitem[Fan et al.(2001)]{Fan01}
Fan, X., et al. 2001, A Survey of z $\sim$ 5.8 Quasars in the Sloan Digital Sky Survey. I. Discovery of Three New Quasars and the Spatial Density of Luminous Quasars at z $\sim$ 6, AJ, 122, 2833

\bibitem[Fan et al.(2002)]{Fan02}
Fan, X., et al. 2002, Evolution of the Ionizing Background and the Epoch of Reionization from the Spectra of $z\sim 6$ Quasars, AJ, 123, 1247


%\bibitem[Fukugita \etal (1996)]{Fukugita96}
%Fukugita, M., et al. 1996, The Sloan Digital Sky Survey Photometric System, AJ, 111, 1748F

%\bibitem[Gnedin \& Fan (2006)]{Gnedin06}
%Gnedin, N, Y., \& Fan, X., 2006, Cosmic Reionization Redux, ApJ, 648, 1G

\bibitem[Haislip et al.(2006)]{Haislip06}
Haislip, J. B., et al. A Photometric Redshift of z = 6.39 $\pm$ 0.12 for GRB 050904, Nature, 440, 181

\bibitem[Hewett et al.(2006)]{Hewett06}
Hewett, P. C., et al. 2006, The UKIRT Infrared Deep Sky Survey $ZY JHK$ Photometric System: Passbands and Synthetic Colours, MNRAS, 367, 454

\bibitem[Hillenbrand et al.(2002)]{Hillenbrand02}
Hillenbrand, L. A., et al. 2002, The $Y$ Band at 1.035 Microns: Photometric Calibration and the DwarfStellar/Substellar Color Sequence, PASP, 114, 708

\bibitem[Hwang et al.(2007)]{Hwang07}
Hwang, N., et al. 2007, An Optical Source Catalog of the North Ecliptic Pole Region, ApJS, 172, 83

\bibitem[Im et al.(2002)]{Im02}
Im, M., et al. 2002, The DEEP Groth Strip Survey. X. Number Density and Luminosity Function of Field E/S0 Galaxies at z $<$ 1, ApJ, 571, 36

\bibitem[Im et al.(2010)]{Im10}
Im, M., et al. 2010, Seoul National University 4K $\times$ 4K Camera(SNUCAM) for Maidanak Observatory, JKAS, 43, 75

\bibitem[Jang et al.(2011)]{Jang11}
Jang, M., et al. 2011, Dust Properties in the Afterglow of GRB 071025 at z $\sim$ 5, ApJ, 741, L20

\bibitem[Jeon et al.(2010)]{Jeon10}
Jeon, Y., et al. 2010, Optical Images and Source Catalog of AKARI North Ecliptic Pole Wide Survey Field,
ApJS, 190, 166

%\bibitem[Johnson (1966)]{Johnson66}
%Johnson, H, L., 1966, Astronomical Measurements in the Infrared, ARAA 4,193

\bibitem[Kang \& Im (2009)]{Kang09}
Kang, E., \& Im, M. 2009, Overdensities of Galaxies at z $\sim$ 3.7 in Chandra Deep Field-South, ApJ, 691, L33

\bibitem[Kashikawa et al.(2004)]{Kashikawa04}
Kashikawa, N., et al. 2004, The Subaru Deep Field: The optical Imaging Data, PASJ, 56, 1011

%\bibitem[Kauffmann \& Haehnelt (2000)]{Kauffmann00}
%Kauffmann, G., \& Haehnelt, M., 2000, A unified model for the evolution of galaxies and quasars, MNRAS, 311, 76K

%\bibitem[Kim \etal (2008)]{Kim08}
%Kim, M., et al. 2008, Decomposition of the Host Galaxies of Active Galactic Nuclei Using Hubble Space Telescope Images, ApJS, 179, 283K

%\bibitem[Koekemoer \etal (2007)]{Koekemoer07}
%Koekemoer, A. M., et al. 2007, The COSMOS Survey: Hubble Space Telescope Advanced Camera for Surveys Observations and Data Processing, ApJS, 172, 196K

\bibitem[Lawrence et al.(2007)]{Lawrence07}
Lawrence, A., et al. 2007, The UKIRT Infrared Deep Sky Survey (UKIDSS), MNRAS, 379, 1599

\bibitem[Lee et al.(2007)]{Lee07}
Lee, H. M., et al. 2007, Nature of Infrared Sources in 11 $\mu m$ Selected Sample from Early Data of the \textit{AKARI} North Ecliptic Pole Deep Survey, PASJ, 59, S529

\bibitem[Lee et al.(2010)]{Lee10}
Lee, I., Im, M., \& Urata, Y., et al. 2010,
First Korean Observations of Gamma-Ray Burst Afterglows at Mt. Lemmon Optical Astronomy Observatory (LOAO),
JKAS, 43, 95

\bibitem[Lilly et al.(1995)]{Lilly95}
Lilly, S. J., et al. 1995, The Canada-France Redshift Survey. VI. Evolution of the Galaxy Luminosity Function to z approximately 1, ApJ, 455, 50

\bibitem[Manduca \& Bell(1979)]{Manduca79}
Manduca, A., \& Bell, R. A. 1979, Atmospheric Extinction in the Near Infrared, PASP, 91, 848

\bibitem[Matsuhara et al.(2006)]{Matsuhara06}
Matsuhara, H., et al. 2006, Deep Extragalactic Surveys around the Ecliptic Poles with \textit{AKARI} (ASTRO-F), PASJ, 58, 673

\bibitem[Monet et al.(2003)]{Monet03}
Monet, J., et al. 2003, The USNO-B Catalog, AJ, 125, 984

%\bibitem[Moon \etal (2008)]{Moon08}
%Moon, B., et al. 2008, KASINICS: The Near Infrared Camera System for the BOAO 1.8m Telescope, PASJ, 60, 849

\bibitem[Mortlock et al.(2011)]{Mortlock11}
Mortlock, D. J., et al. 2011, A Luminous Quasar at a Redshift of $z=7.085$, Nature, 474, 616

\bibitem[Park et al.(2011)]{Park11}
Park, W.-K., Pak, S., Im, M., et al. 2011, PASP, submitted

\bibitem[Perryman et al.(1997)]{Perryman97}
Perryman, M. A. C., et al. 1997, The \textit{HIPPARCOS} Catalogue, A\&A, 323L, 49

\bibitem[Perley et al.(2010)]{Perley10}
Perley, D. A., et al. 2010, Evidence for Supernova-Synthesized Dust from the Rising Afterglow of GRB071025 at z $\sim$ 5, MNRAS, 406, 2473

%\bibitem[Pierre \etal (2004)]{Pierre04}
%Pierre, M., et al. 2004, The XMM-LSS survey. Survey design and first results, JCAP, 09, 011P

%\bibitem[Schlegel \etal (1998)]{Schlegel98}
%Schelegel, D., et al. 1998, Maps of Dust Infrared Emission for Use in Estimation of Reddening and Cosmic Microwave Background Radiation Foregrounds ApJ, 500, 525S

%\bibitem[Selby \etal (1975)]{Selby75}
%Selby, J. E. A., \& McClatchey, R, A., 1975, Atmospheric Transmittance From 0.25 to 28.5: Computer Code LOWTRAN 3, Air Force Cambridge Research Laboratories

\bibitem[Shim et al.(2006)]{Shim06}
Shim, H., et al. 2006, Deep $u*$ and $g$ Band Imaging of the \textit{Spitzer} Space Telescope First Look Survey Field: Observations and Source Catalogs, ApJS, 164, 435

\bibitem[Shim et al.(2007)]{Shim07}
Shim, H., et al. 2007, Massive Lyman Break Galaxies at z$\sim$3 in the \textit{Spitzer} Extragalactic First Look Survey, ApJ, 669, 749

\bibitem[Shimasaku et al.(2005)]{Shimasaku05}
Shimasaku, K., et al. 2005, Number Density of Bright Lyman-Break Galaxies at
 $z \sim 6$ in the Subaru Deep Field, PASJ, 57, 447

\bibitem[Skrutskie et al.(2006)]{Skrutskie06}
Skrutskie, M. F., et al. 2006, The Two Micron All Sky Survey (2MASS), AJ, 131, 1163

\bibitem[Song \& Im(2009)]{Song09}
Song, M., \& Im, M. 2009, AKARI Lightens the 15 $\mu$m Universe at $1<z<1.5$:
 15 μm Observation of the Extended Groth Strip, ASP Conf. Series Vol 418, 527

\bibitem[Steidel et al.(2003)]{Steidel03}
Steidel, C. C., et al. 2003, Lyman Break Galaxies at Redshift z$\sim$3: Survey Description and Full Data Set, ApJ, 592, 728

\bibitem[Vanden Berk et al.(2001)]{Vandenberk01}
Vanden, B. D. E., et al. 2001, Composite Quasar Spectra from the Sloan Digital Sky Survey, AJ, 122, 549

\bibitem[Venemans et al.(2007)]{Venemans07}
Venemans, B. P., et al. 2007, High Redshift QSOs in the UKIDSS Large Area Survey, ASPC, 379, 43

\bibitem[Vogt et al.(2005)]{Vogt05}
Vogt, N. P., et al. 2005, The DEEP Groth Strip Survey. I. The Sample, ApJS, 159, 41

\bibitem[Warren et al.(2007)]{Warren07}
Warren, S. J., et al. 2007, The United Kingdom Infrared Telescope Infrared Deep Sky Survey First Data Release, MNRAS, 375, 213

\bibitem[Willott et al.(2007)]{Willott07}
Willott, C. J., et al. 2007, Four Quasars above Redshift 6 Discovered by the Canada-France High-z Quasar Survey,  AJ, 134, 2435

\end{thebibliography}
\end{document}